\documentclass{ws-ijqi}%
\usepackage{amsfonts}
\usepackage{amsmath}
\usepackage{amssymb}
\usepackage{graphicx}
\usepackage{url}%
\begin{document}

\markboth{Mark M. Wilde, Federico Spedalieri, Jonathan P. Dowling, Hwang Lee}
{Alternate Scheme for Optical Cluster-State Generation without Number-Resolving Photon Detectors}

\title{ALTERNATE SCHEME FOR OPTICAL CLUSTER-STATE GENERATION WITHOUT NUMBER-RESOLVING
PHOTON DETECTORS}

\author{MARK M. WILDE}

\address{Communication Sciences Institute, University of Southern California,\\
Los Angeles, California 90089, United States\\
mark.wilde@usc.edu}

\author{FEDERICO SPEDALIERI}

\address{Department of Electrical Engineering, University of California, Los
Angeles,\\
Los Angeles, California 90095, United States\\
federico@ee.ucla.edu}

\author{JONATHAN P. DOWLING}
\author{HWANG LEE}

\address{Hearne Institute for Theoretical Physics, Department of Physics and
Astronomy, Louisiana State University,\\
Baton Rouge, Louisiana 70803-4001, United States\\
jdowling@phys.lsu.edu, hwlee@phys.lsu.edu}

\maketitle

\begin{history}
\received{Day Month Year}
\revised{Day Month Year}
\end{history}

\begin{abstract}
We design a controlled-phase gate for linear optical quantum computing by
using photodetectors that cannot resolve photon number. An intrinsic
error-correction circuit corrects errors introduced by the detectors. Our
controlled-phase gate has a 1/4 success probability. Recent development in
cluster-state quantum computing has shown that a two-qubit gate with non-zero
success probability can build an arbitrarily large cluster state with only
polynomial overhead. Hence, it is possible to generate optical cluster states
without number-resolving detectors and with polynomial overhead.
\end{abstract}


\section{Introduction}

The protocol of one-way quantum computing opened a new paradigm in quantum
computation. Single-qubit measurements on a set of highly entangled
qubits---the so called cluster state\cite{prl2001raus}---perform the
computation in one-way quantum computing. On the other hand, in the usual
quantum-circuit approach, Knill, Laflamme, and Milburn (KLM) utilized
the idea of gate teleportation\cite{nature1999got} for scalable optical quantum computing.
 KLM introduced a non-deterministic
two-qubit controlled-phase gate that succeeds with probability arbitrarily
close to one, given a sufficiently large ancilla state\cite{nat2001klm}. Nielsen later applied
the KLM method to one-way quantum computing in the effort to decrease the
number of required physical resources\cite{prl2004niel}. \textquotedblleft
Qubit fusion\textquotedblright\cite{prl2005browne} and \textquotedblleft
quantum parity check\textquotedblright\ operations\cite{pra2001franson} have
improved the scalability of linear optical quantum computation (see Ref. \refcite{kok:135}\ for a review).
Two recent experiments have implemented the one-way quantum computing
model with photonics\cite{nature2005zeilinger,natphys2007pan}.

Many researchers consider the number-resolving-detector requirement as
inevitable for either the one-way or quantum circuit approach for linear
optical quantum computing. The number-resolving capability is an essential
requirement in a quantum repeater protocol for long distance communication\cite{nature2001duan}. Browne and Rudolph showed that number-resolving
detectors are not necessary for building optical cluster states by using
redundant encoding in Type-II\ fusion\cite{prl2005browne}. We address the
necessity of number-resolving photon detectors for the implementation of
linear optical quantum computing. We provide an alternate scheme for
building\ cluster states using an optical controlled-phase gate that does not
rely on number-resolving detectors. We note that some authors recently
presented another scheme for building cluster states without number-resolving
detectors\cite{arxiv2007varn}.

We utilize the KLM-type ancilla state of four qubits with a source of pure
polarization-entangled Bell states. We adopt a
polarization-encoding scheme\cite{pra2006sped}. It has the advantage of
immunity to photon loss errors\cite{prl2005ralph}. Our
scheme requires a source of high-fidelity
polarization-entangled Bell states\cite{nat2006schields}. The
controlled-phase gate succeeds with probability $1/4$ with the aid of a
four-qubit ancilla state. Intriguingly, previous researchers have suggested
using a probabilistic two-qubit gate with non-zero success probability to
build an arbitrarily large cluster state\cite{pra2005kok,prl2005duan}.
Preparation of two-dimensional cluster states requires only polynomial
overhead\cite{arxiv2006kieling,pra2006kieling}. Our controlled-phase gate
scheme removes the necessity of number-resolving detectors for building an
arbitrarily large optical cluster state.

\begin{figure}
[ptb]
\begin{center}
\includegraphics[
height=1.9943in,
width=3.2439in
]%
{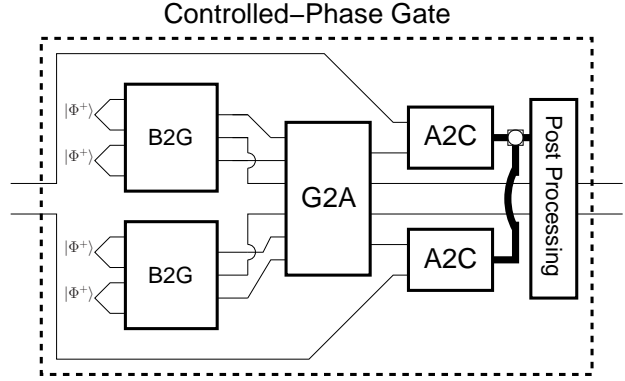}%
\caption{Our implementation of a controlled-phase gate. B2G is the
Bell-to-GHZ\ converter, G2A is the GHZ-to-four-qubit-Ancilla converter, and
A2C is the four-qubit-Ancilla-to-Controlled-phase converter.}%
\label{fig_C-Z}%
\end{center}
\end{figure}

Figure~\ref{fig_C-Z} outlines our controlled-phase-gate design.%
B2G (Bell-to-GHZ) converts Bell states to GHZ\ states with $1/2$ success
probability. G2A (GHZ-to-four-qubit-Ancilla) corrects for possible errors
introduced in B2G. G2A converts pure GHZ\ states to the following four-qubit
ancilla state $\left\vert t_{1}^{\prime}\right\rangle $\cite{nature1999got,nat2001klm,pra2006sped}:%
\begin{equation}
\left(  \left\vert HVVH\right\rangle +\left\vert VHVH\right\rangle +\left\vert
VHHV\right\rangle -\left\vert HVHV\right\rangle \right)  /2
\end{equation}
The conversion probability is $1/2$ given two pure GHZ\ states. A2C
(four-qubit-Ancilla-to-Controlled-phase)\ uses $\left\vert t_{1}^{\prime
}\right\rangle $ to perform a controlled-phase gate with $1/4$ success
probability. The gate heralds success using photon detectors that do not
resolve photon number.
\begin{figure}[tbp] \centering
\begin{tabular}
[c]{cc}%
{\parbox[b]{2.1417in}{\begin{center}
\includegraphics[
height=2.002in,
width=2.1417in
]%
{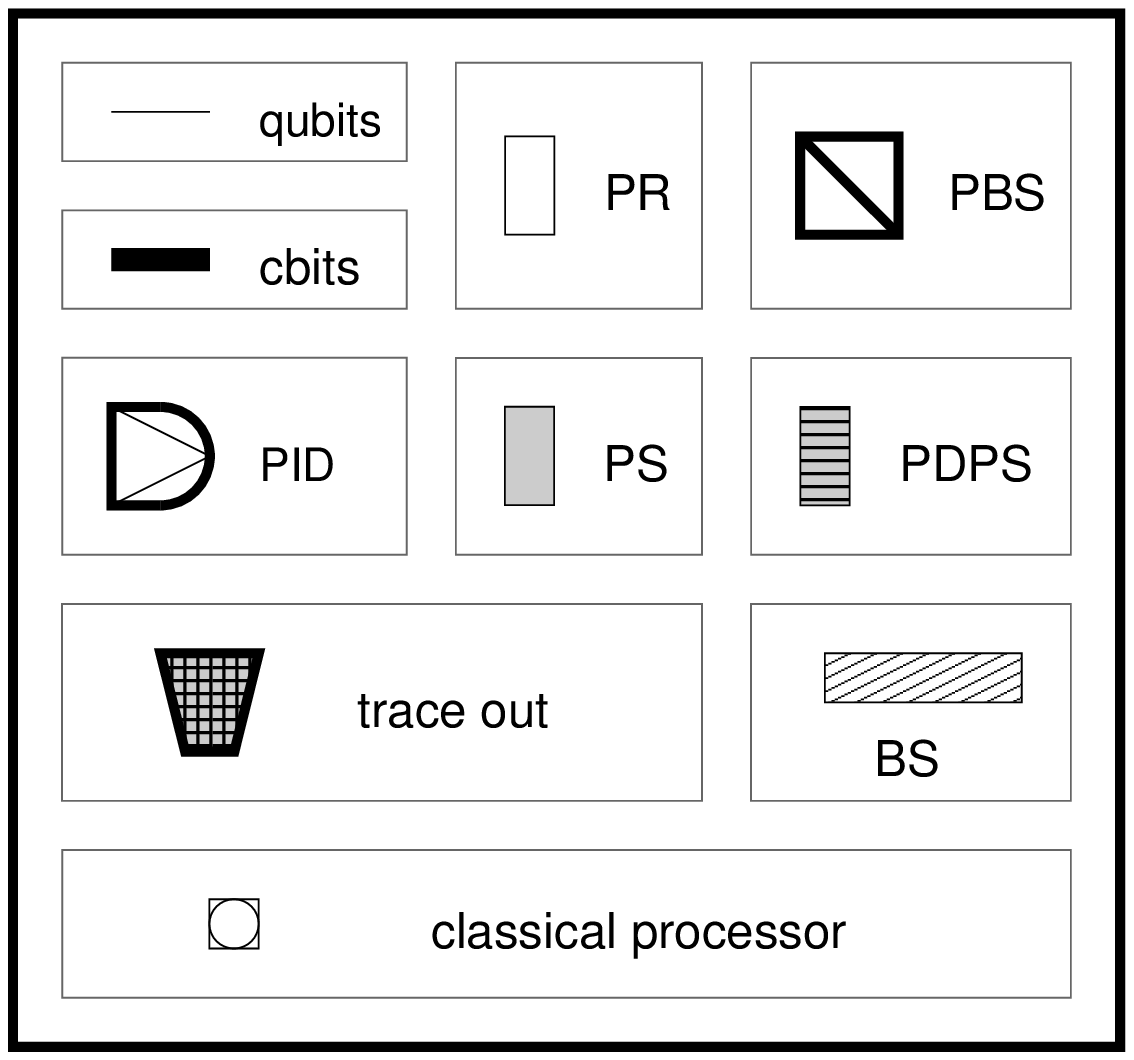}%
\\
(a)
\end{center}}}%
&
{\parbox[b]{2.0107in}{\begin{center}
\includegraphics[
height=1.7604in,
width=2.0107in
]%
{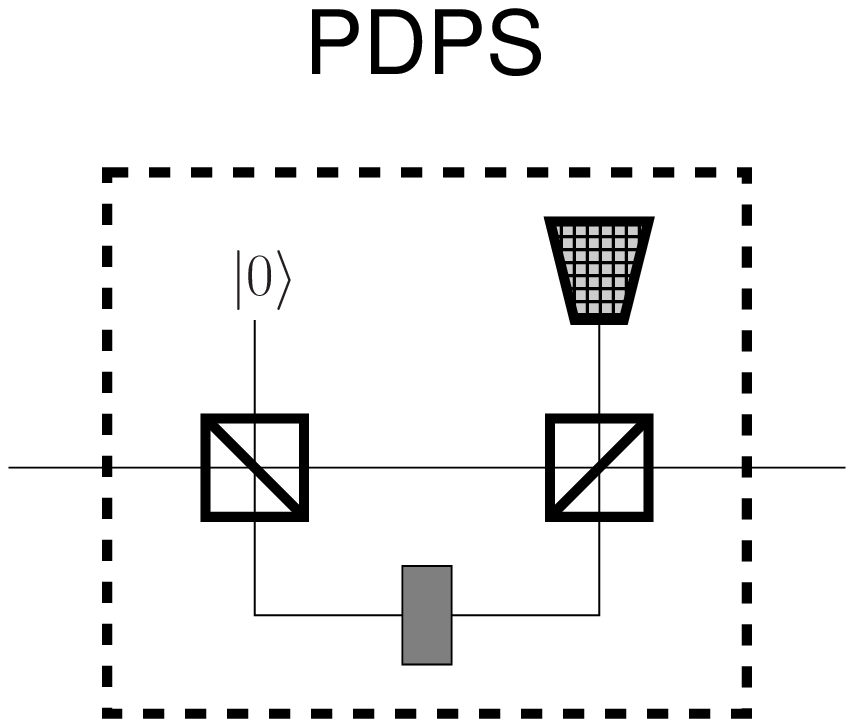}%
\\
{}(b)
\end{center}}}%
\\%
{\parbox[b]{1.8822in}{\begin{center}
\includegraphics[
height=1.4308in,
width=1.8822in
]%
{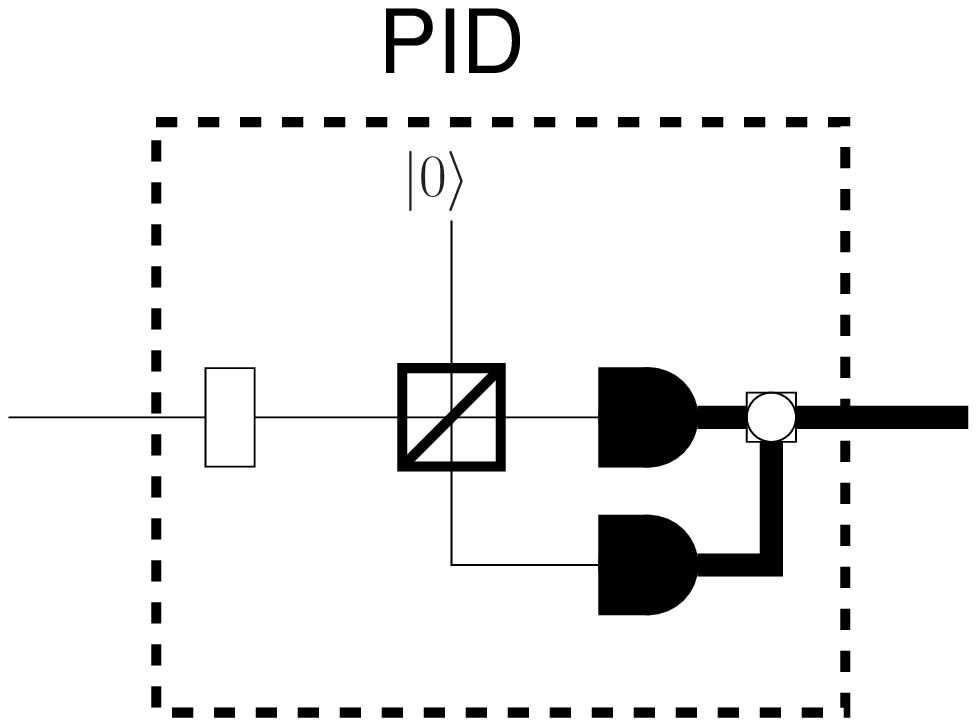}%
\\
{}(c)
\end{center}}}%
&
\hspace{.15in}%
\raisebox{-0.1505in}{\parbox[b]{2.6748in}{\begin{center}
\includegraphics[
height=1.7565in,
width=2.6748in
]%
{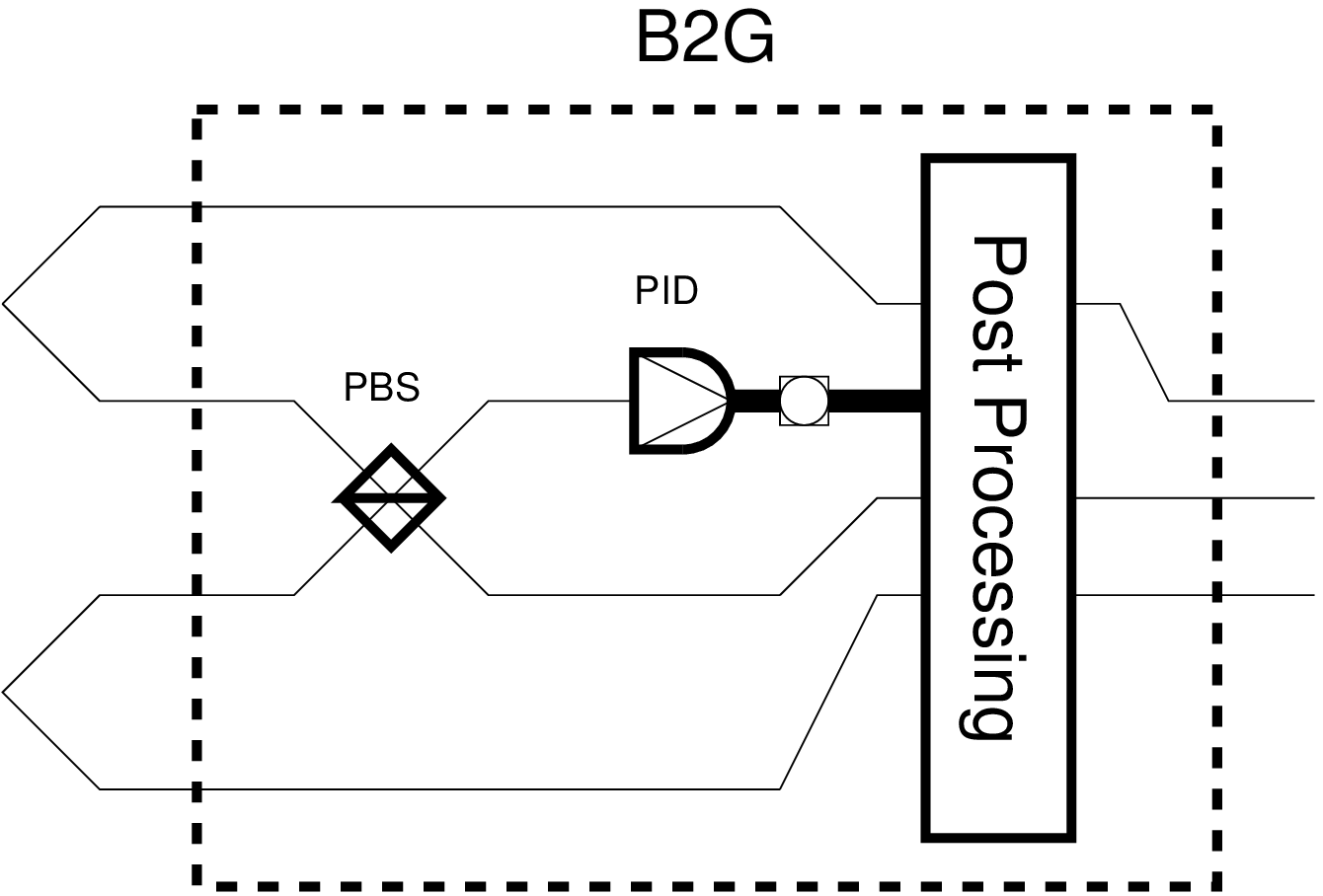}%
\\
(d)
\end{center}}}%
\end{tabular}%
\caption
{In the legend (a), thin lines denote qubits  and thick lines denote classical
bits. A classical processor performs post processing based on measurement results. The trash symbol denotes \textquotedblleft
tracing
out\textquotedblright\ a mode. A polarization-independent
detector (PID) detects photons independent of polarization. A
polarization rotator (PR) rotates the polarization basis by $\frac{\pi}{4}$.
A phase shifter (PS) rotates the global
phase. A polarizing
beam splitter (PBS) transmits horizontally-polarized photons and reflects
vertically-polarized photons. A polarization-dependent phase shifter (PDPS)
shifts the phase of vertically-polarized photons only (Figure
\ref{fig_PDPS_PID}(b)). The last linear optical element in (a)
is the beam splitter\ (BS). A polarization-dependent phase shifter (b).
A polarization-independent detector (c).
B2G (d) is the Bell state to GHZ state converter.}%
\label{fig_PDPS_PID}%
\end{figure}%

Figure~\ref{fig_PDPS_PID}(a) provides a guide to all linear optical elements.
Figure~\ref{fig_PDPS_PID}(b) shows a polarization-dependent phase shifter and
Figure~\ref{fig_PDPS_PID}(c) a polarization-independent detector.

\section{Polarization-Independent Detectors}

Polarization-independent detectors (PIDs) detect the number of photons in a
spatial mode independent of their polarization (Figure~\ref{fig_PDPS_PID}(c)).
Two ordinary photon detectors follow the polarization rotator (PR) and the
polarizing beam splitters (PBS). The classical processor processes the result
of the two detectors' photon number measurement. It sends classical control
signals to optical elements that perform conditional post-processing on the
remaining computational modes. We illustrate its operation on the
computational basis $\{\left\vert H\right\rangle ,\left\vert V\right\rangle
\}$ of one incoming spatial mode:
\begin{eqnarray}
\left\vert H\right\rangle
\ \ & \underrightarrow{\text{PID}}\ \ &  \left(  \left\vert H0\right\rangle
+\left\vert 0V\right\rangle \right)  /\sqrt{2} \\
\left\vert V\right\rangle
\ \ & \underrightarrow{\text{PID}}\ & \left(  -\left\vert H0\right\rangle
+\left\vert 0V\right\rangle \right)  /\sqrt{2}
\end{eqnarray}
The outcome of the photon
number measurement in each case is either $\left\vert H0\right\rangle $ or
$\left\vert 0V\right\rangle $\ with probability $1/2$. The detectors only gain
information about the photon number in the incoming spatial mode---they learn
nothing about polarization.

We illustrate a useful feature of a PID. Suppose we have a
polarization-entangled state
\begin{equation}
\left\vert \Phi_{d}^{+}\right\rangle
\equiv(\left\vert H\right\rangle ^{\otimes d}+\left\vert V\right\rangle
^{\otimes d})/\sqrt{2}
\end{equation}
We can measure the photon number of the last mode of
$\left\vert \Phi_{d}^{+}\right\rangle $ using a PID and obtain the state
$\left\vert \Phi_{d-1}^{+}\right\rangle $. This retaining of entanglement is
not possible with a typical photon number measurement.

A PID processing $\left\vert \Phi_{d}^{+}\right\rangle $'s last mode gives the
state:%
\begin{equation}
\left\vert \Phi_{d}^{+}\right\rangle \ \ \underrightarrow{\text{PID}%
}\ \ 2^{-1/2}\left[  \left\vert \Phi_{d-1}^{-}\right\rangle \otimes\left\vert
H0\right\rangle +\left\vert \Phi_{d-1}^{+}\right\rangle \otimes\left\vert
0V\right\rangle \right]
\end{equation}
The state after measurement is the polarization-entangled state $\left\vert
\Phi_{d-1}^{-}\right\rangle $ if we measure $\left\vert H0\right\rangle $ and
state $\left\vert \Phi_{d-1}^{+}\right\rangle $ if we measure $\left\vert
0V\right\rangle $. The classical processor in the PID\ in
Figure~\ref{fig_PDPS_PID}(c) obtains the result of the measurement and forwards
it to a set of post-processing linear optical elements. Post-processing is as
follows: do nothing for measurement result $\left\vert 0V\right\rangle $, or
evolve state $\left\vert \Phi_{d-1}^{-}\right\rangle $ to $\left\vert
\Phi_{d-1}^{+}\right\rangle $ for measurement result $\left\vert
H0\right\rangle $. Evolve by performing a PDPS\ of $\pi$ on exactly one mode
of $\left\vert \Phi_{d-1}^{-}\right\rangle $. So we possess $\left\vert
\Phi_{d-1}^{+}\right\rangle $ as a resource for subsequent computations. We
exploit this feature in B2G.

\section{Preparation of a GHZ state from two Bell states}

Suppose we have a source of pure Bell states
\begin{equation}
\left\vert \Phi^{+}\right\rangle
\equiv(\left\vert HH\right\rangle +\left\vert VV\right\rangle )/\sqrt{2}
\end{equation}
Note the following conventions:
\begin{eqnarray}
\left\vert \text{GHZ}^{+}\right\rangle
& \equiv & (\left\vert HHH\right\rangle +\left\vert VVV\right\rangle )/\sqrt{2} \\
\left\vert \text{GHZ}^{-}\right\rangle & \equiv & (\left\vert HHH\right\rangle
-\left\vert VVV\right\rangle )/\sqrt{2}
\end{eqnarray}
B2G in Figure~\ref{fig_PDPS_PID}(d)
converts pure Bell states to a mixture of $\left\vert \text{GHZ}%
^{+}\right\rangle $ and $\left\vert V0H\right\rangle $. B2G is nothing but the
quantum parity check\cite{pra2001franson} and similar to Type I fusion\cite{prl2005browne}. A mixture results because the photon detectors are not
number resolving. We use polarization-independent detectors in our scheme
where the scheme in Ref. \refcite{prl2005browne} uses detectors dependent on the
polarization of the incoming photons.

B2G begins by feeding in two Bell states at the four input ports in
Figure~\ref{fig_PDPS_PID}(d). $\left\vert \Phi^{+}\right\rangle \left\vert
\Phi^{+}\right\rangle $\ propagates after the PBS as follows (with the third
mode rearranged as the last mode):%
\begin{multline}
\left\vert \Phi^{+}\right\rangle \left\vert \Phi^{+}\right\rangle
\ \ \underrightarrow{\text{PBS}}\ \ \left(  \left\vert HHH\right\rangle
\left\vert H\right\rangle +\left\vert VVV\right\rangle \left\vert
V\right\rangle \right)  /2\\
+\left(  \left\vert V0H\right\rangle \left\vert \left(  H,V\right)
\right\rangle +\left\vert H\left(  H,V\right)  V\right\rangle \left\vert
0\right\rangle \right)  /2
\end{multline}
We perform a PID on the last mode. The quantum state becomes the following
before the detectors in the PID:%
\begin{multline}
\left(  \left\vert \text{GHZ}^{-}\right\rangle \left\vert H0\right\rangle
+\left\vert \text{GHZ}^{+}\right\rangle \left\vert 0V\right\rangle \right)
/\sqrt{2}\ -\label{eqn_B2G_q_state}\\
\left\vert V0H\right\rangle \left(  \left\vert H^{2}0\right\rangle -\left\vert
0V^{2}\right\rangle \right)  /\sqrt{8}+\left\vert H\left(  HV\right)
V\right\rangle \left\vert 00\right\rangle /2
\end{multline}
$\left\vert H^{2}\right\rangle $ denotes two horizontally polarized photons in
a given path. The detectors in the PID measure the last two modes and\ cannot
distinguish between $\left\vert H0\right\rangle $ and $\left\vert
H^{2}0\right\rangle $ or $\left\vert 0V\right\rangle $ and $\left\vert
0V^{2}\right\rangle $ because they cannot resolve photon number. Thus
$\left\vert H0\right\rangle $ and $\left\vert H^{2}0\right\rangle $ refer to
the same measurement result so we name them $\left\vert H^{n}0\right\rangle $
and $\left\vert 0V^{n}\right\rangle $ where $n$ is an arbitrary positive
integer. Post-processing on the quantum state in (\ref{eqn_B2G_q_state})
is as follows: discard and start over if we measure $\left\vert
00\right\rangle $, perform a PDPS\ of $\pi$ on the first mode if we measure
$\left\vert H^{n}0\right\rangle $, or do nothing if we measure $\left\vert
0V^{n}\right\rangle $.

The state becomes the mixture $\rho_{\text{PGHZ}}$\ (the partial GHZ\ mixture)
after the above conditioning.%
\begin{equation}
\rho_{\text{PGHZ}}=\left(  2\left\vert \text{GHZ}^{+}\right\rangle
\left\langle \text{GHZ}^{+}\right\vert +\left\vert V0H\right\rangle
\left\langle V0H\right\vert \right)  /3 \label{eqn_mixture}%
\end{equation}
We obtain a pure GHZ\ state with probability $1/2$ after performing B2G on two
pure Bell states.%

\begin{figure}[tbp] \centering
\begin{tabular}
[c]{cc}%
{\parbox[b]{1.967in}{\begin{center}
\includegraphics[
height=1.391in,
width=1.967in
]%
{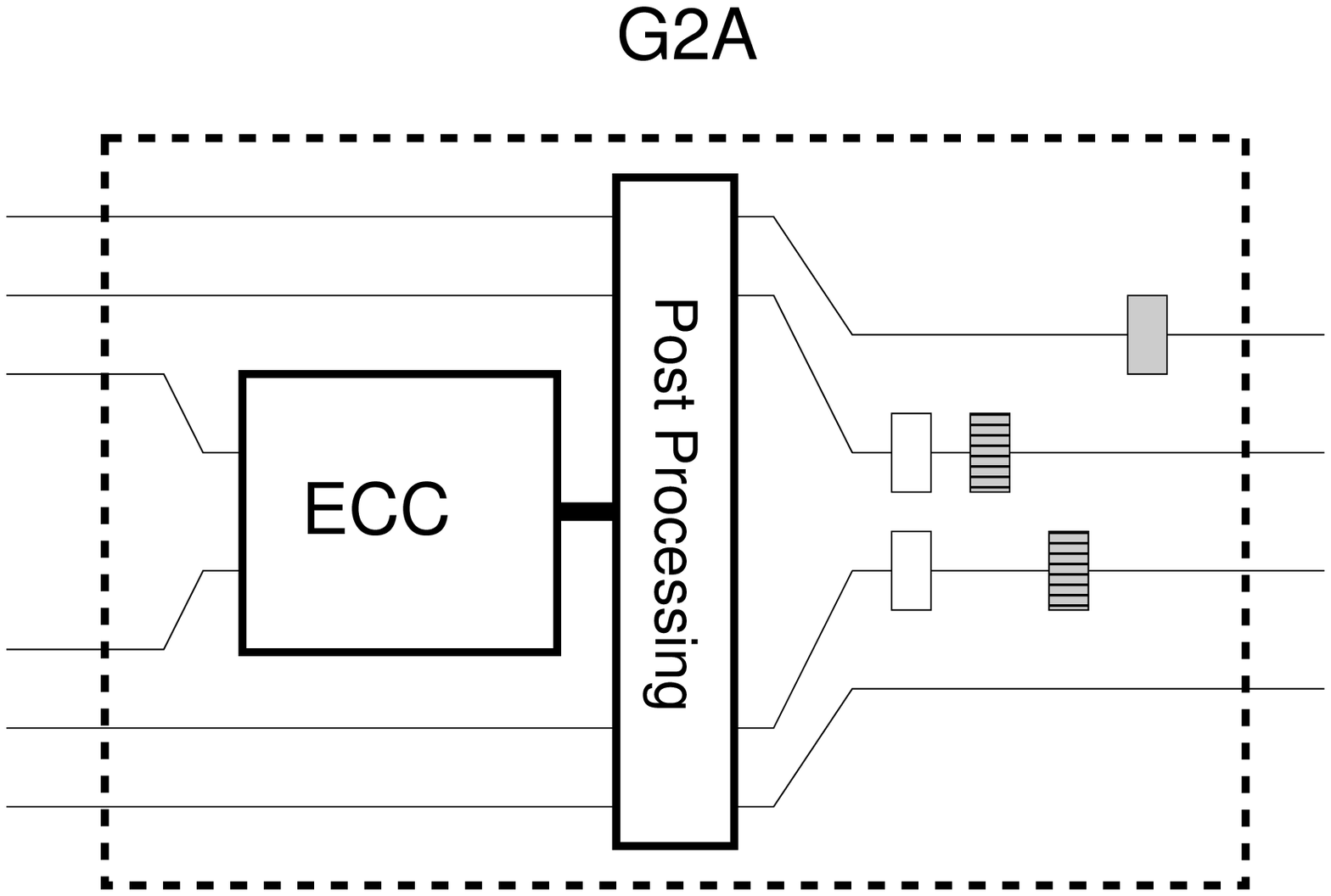}%
\\
{}(a)
\end{center}}}
&
\hspace{.25in}%
{\parbox[b]{2.27in}{\begin{center}
\includegraphics[
height=1.223in,
width=2.27in
]%
{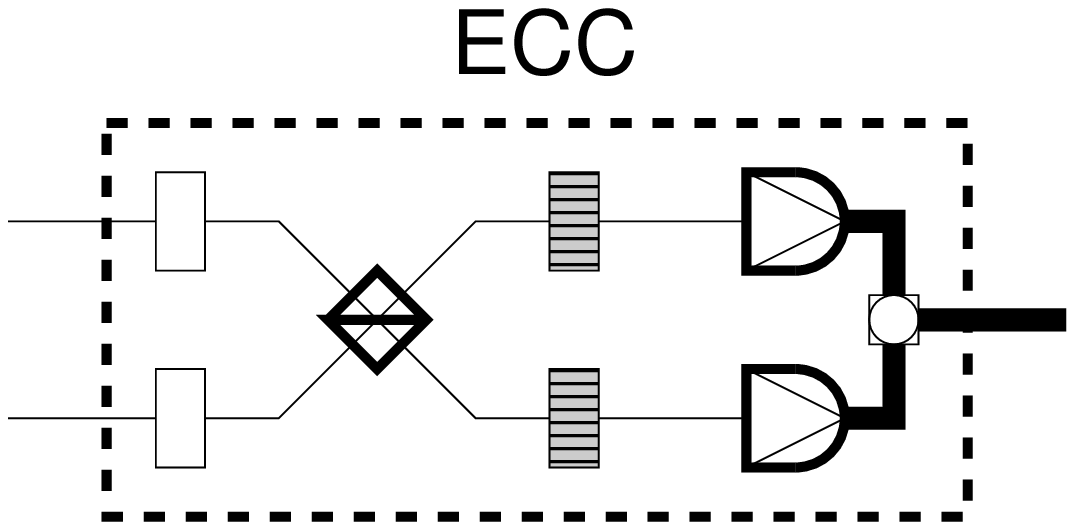}%
\\
{}(b)
\end{center}}}
\end{tabular}%
\caption{G2A (a) converts the mixture $\rho_{\text{PGHZ}}$ in Eqn. (\ref
{eqn_mixture})
to
$\left\vert t_{1}^{\prime}\right\rangle$. It
includes error correcting in ECC followed by
post-processing to generate $\left\vert t_{1}^{\prime}\right\rangle
$. ECC (b) corrects for errors from B2G. Each
PDPS introduces a relative phase of $\frac{\pi}{4}$.}%
\label{fig_G2A_ECC}%
\end{figure}%

\section{Preparation of the four-qubit ancilla state from two GHZ states}

G2A in Figure~\ref{fig_G2A_ECC}(a) converts the output of two parallel runs of
B2G to the four-qubit ancilla state $\left\vert t_{1}^{\prime}\right\rangle $.
G2A performs this conversion with a success probability of $1/2$ when pure
GHZ\ states are input. We perform this conversion using photon detectors which
are not number resolving. We have a procedure that corrects for the error
introduced into the mixed state $\rho_{\text{PGHZ}}$ in (\ref{eqn_mixture}%
). This conversion only has a non-unit probability of success, but we know
whether the conversion fails or succeeds. Two parallel runs of B2G generate
two copies of $\rho_{\text{PGHZ}}$. We input four Bell states to the two B2Gs.

\subsection{Intrinsic Error Correction Circuit}

The second mode of $\rho
_{\text{PGHZ}}$ may have an error. So we send the second mode of each copy of
$\rho_{\text{PGHZ}}$ through the ECC (Error Correction Circuit) depicted in
Figure~\ref{fig_G2A_ECC}(b). ECC is the first part of G2A. ECC\ is similar to
Type II fusion\cite{prl2005browne}. We use it for two purposes.\ It detects
whether two pure GHZ\ states are actually input to G2A. ECC then produces a
state which deterministically converts to the four-qubit ancilla state
$\left\vert t_{1}^{\prime}\right\rangle $. It only produces the convertible
state if two pure GHZ states are at the input of G2A. G2A is similar to a
Type-II fusion gate, which (in the language of Ref. \refcite{prl2005browne}) creates a
\textquotedblleft redundantly-encoded\textquotedblright\ two-qubit cluster
state (across four photons) and simultaneously filters out the unwanted
two-photon parts of the wavefunction of the input states (as described by
Ralph et al. in Ref. \refcite{prl2005ralph}). 

Suppose that the state $\left\vert V0H\right\rangle \left\vert
V0H\right\rangle $ is input to G2A. We can detect this state uniquely by
noting that no one of the four detectors in the two PIDs fire. We discard and
start over if we detect zero photons.

Suppose that either the state $\left\vert \text{GHZ}^{+}\right\rangle
\left\vert V0H\right\rangle $ or $\left\vert V0H\right\rangle \left\vert
\text{GHZ}^{+}\right\rangle $\ is input to ECC. The left column of Table
\ref{tbl_Error_Crct}\ gives possible states of the two middle modes in either
of the two superpositions: $\left\vert \text{GHZ}^{+}\right\rangle \left\vert
V0H\right\rangle $ or $\left\vert V0H\right\rangle \left\vert \text{GHZ}%
^{+}\right\rangle $. Table \ref{tbl_Error_Crct} aids in determining the
resulting state before the detector measurement. Each measurement result has
zero photons in exactly three of the four modes. We discard and start over if
we detect zero photons in exactly three modes.%
\begin{table}
\tbl{The left column contains possible initial states of the two middle modes of
state $\left\vert\mbox{GHZ}\right\rangle \left\vert V0H \right\rangle$ or
state $ \left\vert V0H \right\rangle\left\vert\mbox{GHZ}\right\rangle$.
The right column contains resulting states after ECC processes
initial states. The two PIDs transform the two modes in the initial states
to four modes (see Figure 2(b)).}
{\begin{tabular}
[c]{l|l}\hline\hline
\textbf{Init.} & \textbf{Resulting States}\\\hline
$\left\vert 0H\right\rangle $ & $\left(  \left\vert 0H00\right\rangle
-e^{i\pi/4}\left\vert H000\right\rangle +\left\vert 000V\right\rangle
+e^{i\pi/4}\left\vert 00V0\right\rangle \right)  /2$\\\hline
$\left\vert 0V\right\rangle $ & $\left(  -\left\vert 0H00\right\rangle
-e^{i\pi/4}\left\vert H000\right\rangle -\left\vert 000V\right\rangle
+e^{i\pi/4}\left\vert 00V0\right\rangle \right)  /2$\\\hline
$\left\vert H0\right\rangle $ & $\left(  \left\vert H000\right\rangle
-e^{i\pi/4}\left\vert 0H00\right\rangle +\left\vert 00V0\right\rangle
+e^{i\pi/4}\left\vert 000V\right\rangle \right)  /2$\\\hline
$\left\vert V0\right\rangle $ & $\left(  -\left\vert H000\right\rangle
-e^{i\pi/4}\left\vert 0H00\right\rangle -\left\vert 00V0\right\rangle
+e^{i\pi/4}\left\vert 000V\right\rangle \right)  /2$\\\hline\hline
\end{tabular}}
\label{tbl_Error_Crct}%
\end{table}%

Suppose that two pure GHZ\ states $\left\vert \text{GHZ}^{+}\right\rangle
\left\vert \text{GHZ}^{+}\right\rangle $ are input to ECC. The left column of
Table \ref{tbl_Error_Crct_2} gives possible states of the two middle modes in
the superposition $\left\vert \text{GHZ}^{+}\right\rangle \left\vert
\text{GHZ}^{+}\right\rangle $\ just before the detectors in the two PIDs.%
\begin{table}
\tbl{The left column contains the possible initial states of the two middle modes of
$\left\vert\mbox{GHZ}\right\rangle \left\vert\mbox{GHZ}\right\rangle$.
The right column contains the resulting states after the ECC operation processes
the initial states and just before detection.
We employ the following shorthand notation: $\left\vert 0H0V\right\rangle \equiv\left\vert 1\right\rangle $,
$\left\vert H0V0\right\rangle \equiv\left\vert 2\right\rangle $, $\left\vert
H00V\right\rangle \equiv\left\vert 3\right\rangle $, $\left\vert
0HV0\right\rangle \equiv\left\vert 4\right\rangle $, $\left\vert
HH00\right\rangle \equiv\left\vert 5\right\rangle $, $\left\vert
00VV\right\rangle \equiv\left\vert 6\right\rangle $, $\left\vert
H^{2}000\right\rangle \equiv\left\vert 7\right\rangle $, $\left\vert
0H^{2}00\right\rangle \equiv\left\vert 8\right\rangle $, $\left\vert
00V^{2}0\right\rangle \equiv\left\vert 9\right\rangle $, $\left\vert
000V^{2}\right\rangle \equiv\left\vert 10\right\rangle $.
All resulting states require normalization.}
{\begin{tabular}
[c]{l|l}\hline\hline
\textbf{Init.} & \textbf{Resulting States}\\\hline
$\left\vert HH\right\rangle $ & $e^{\frac{i\pi}{4}}\left(  \left\vert
5\right\rangle -i\left\vert 3\right\rangle -i\left\vert 4\right\rangle
+\left\vert 6\right\rangle -\left\vert 7\right\rangle +\left\vert
9\right\rangle -\left\vert 8\right\rangle +\left\vert 10\right\rangle \right)
$\\\hline
$\left\vert HV\right\rangle $ & $ie^{\frac{i3\pi}{4}}\left(  -i\left\vert
5\right\rangle -\left\vert 3\right\rangle -\left\vert 4\right\rangle
-i\left\vert 6\right\rangle +\left\vert 7\right\rangle -\left\vert
9\right\rangle -\left\vert 8\right\rangle +\left\vert 10\right\rangle \right)
$\\\hline
$\left\vert VH\right\rangle $ & $ie^{\frac{i3\pi}{4}}\left(  -i\left\vert
5\right\rangle -\left\vert 3\right\rangle -\left\vert 4\right\rangle
-i\left\vert 6\right\rangle -\left\vert 7\right\rangle +\left\vert
9\right\rangle +\left\vert 8\right\rangle -\left\vert 10\right\rangle \right)
$\\\hline
$\left\vert VV\right\rangle $ & $e^{\frac{i\pi}{4}}\left(  \left\vert
5\right\rangle -i\left\vert 3\right\rangle -i\left\vert 4\right\rangle
+\left\vert 6\right\rangle +\left\vert 7\right\rangle +\left\vert
9\right\rangle -\left\vert 8\right\rangle -\left\vert 10\right\rangle \right)
$\\\hline\hline
\end{tabular}}
\label{tbl_Error_Crct_2}%
\end{table}
We determine a method to correct for the error introduced in B2G by analyzing
Table \ref{tbl_Error_Crct_2}. We discard the computation if we measure zero
photons in exactly three modes---the resulting states $\left\vert
7\right\rangle ,\left\vert 8\right\rangle ,\left\vert 9\right\rangle
,\left\vert 10\right\rangle $ are in Table \ref{tbl_Error_Crct_2}, as well as all the resulting
states given in Table \ref{tbl_Error_Crct}. \textit{We only need detectors that distinguish
between no photons and some photons}. Keep the state if zero photons are in
only two modes. We thus determine with certainty whether two pure GHZ\ states
are input to G2A.

We analyze the result of ECC only when two pure GHZ\ states $\left\vert
\text{GHZ}^{+}\right\rangle \left\vert \text{GHZ}^{+}\right\rangle $ are input
to G2A. Reorder the state $\left\vert \text{GHZ}^{+}\right\rangle \left\vert
\text{GHZ}^{+}\right\rangle $ so that the middle two modes become the last two
modes: $(\left\vert HHHH\right\rangle \left\vert HH\right\rangle +\left\vert
HHVV\right\rangle \left\vert HV\right\rangle +\left\vert VVHH\right\rangle
\left\vert VH\right\rangle +\left\vert VVVV\right\rangle \left\vert
VV\right\rangle )/2$. Table \ref{tbl_Measure_Results}\ entries give the
resulting state after ECC.

ECC (Figure~\ref{fig_G2A_ECC}(b)) works only if the original Bell states are
pure. If the original Bell states have timing jitters, a PBS can purify them.
If, on the other hand, the Bell-state source has the possibility of emitting a
four-photon pair, the purification is seemingly not possible. Therefore, our
scheme doesn't not work if we use spontaneous parametric down conversion as
our source of the Bell pairs. It is possible to implement the desired ECC if
we use the source without multiple-pair emission such as the recently
developed semiconductor source\cite{nat2006schields}.%

\begin{table}
\tbl{The left column contains measurement
results when passing the two middle modes of
$\left\vert\mbox{GHZ}\right\rangle \left\vert\mbox{GHZ}\right\rangle$
through ECC. The right column contains the resulting states
of the four other modes which do not pass through ECC.}
{\begin{tabular}
[c]{l|l}\hline\hline
\textbf{Meas.} & \textbf{Resulting States}\\\hline
$\left\vert 5\right\rangle ,\left\vert 6\right\rangle $ & $e^{\frac{i\pi}{4}%
}\left(
\begin{array}
[c]{c}%
\left\vert HHHH\right\rangle +i\left\vert HHVV\right\rangle \\
+i\left\vert VVHH\right\rangle +\left\vert VVVV\right\rangle
\end{array}
\right)  /2$\\\hline
$\left\vert 3\right\rangle ,\left\vert 4\right\rangle $ & $e^{-\frac{i\pi}{4}%
}\left(
\begin{array}
[c]{c}%
\left\vert HHHH\right\rangle -i\left\vert HHVV\right\rangle \\
-i\left\vert VVHH\right\rangle +\left\vert VVVV\right\rangle
\end{array}
\right)  /2$\\\hline\hline
\end{tabular}}
\label{tbl_Measure_Results}%
\end{table}%
\begin{table}
\tbl{The left column contains the
measurement results after two A2Cs.
The right column contains post-processing
for the two remaining computational
modes given the measurement result (using the shorthand in Table \ref{tbl_Error_Crct_2}).}
{\begin{tabular}
[c]{l|l}\hline\hline
\textbf{Meas.} & \textbf{Post-processing Operations}\\\hline
$\left\vert 11\right\rangle ,\left\vert 22\right\rangle $ & $\pi$ PR on both
modes, PDPS of $\pi$ on mode 2\\\hline
$\left\vert 12\right\rangle ,\left\vert 21\right\rangle $ & Same as above with
a PS of $\pi$\\\hline
$\left\vert 31\right\rangle ,\left\vert 42\right\rangle $ & $\pi$ PR on mode
2, $\pi$ PS, PDPS of $\pi$ on mode 2\\\hline
$\left\vert 41\right\rangle ,\left\vert 32\right\rangle $ & Same as above with
a PS of $\pi$\\\hline
$\left\vert 34\right\rangle ,\left\vert 43\right\rangle $ & PDPS of $\pi$ on
mode 2\\\hline
$\left\vert 33\right\rangle ,\left\vert 44\right\rangle $ & Same as above with
a PS of $\pi$\\\hline
$\left\vert 14\right\rangle ,\left\vert 23\right\rangle $ & $\pi$ PR on mode
1, $\pi$ PS, PDPS of $\pi$ on mode 2\\\hline
$\left\vert 13\right\rangle ,\left\vert 24\right\rangle $ & Same as above with
a PS of $\pi$\\\hline\hline
\end{tabular}}
\label{tbl_A2C post_proc}%
\end{table}%
\begin{figure}
[pt]
\begin{center}
\includegraphics[
height=1.0131in,
width=1.6851in
]%
{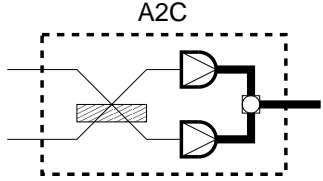}%
\caption{A2C consists of a 50:50 BS followed by two PIDs.}%
\label{fig_A2C}%
\end{center}
\end{figure}

\subsection{Feed-forward to the four-qubit ancilla state}

We perform the following post-processing conditioning on the state resulting
from ECC: perform a PDPS by $\pi$ on modes two and three if we measure either
$\left\vert HH00\right\rangle $ or $\left\vert 00VV\right\rangle $, or perform
a PS\ by $\pi/2$ if we measure either $\left\vert H00V\right\rangle $ or
$\left\vert 0HV0\right\rangle $. The resulting (unnormalized) state is as
follows: $\left\vert HHHH\right\rangle -i\left\vert HHVV\right\rangle
-i\left\vert VVHH\right\rangle +\left\vert VVVV\right\rangle $. The later
elements of G2A convert the above state deterministically to $\left\vert
t_{1}^{\prime}\right\rangle $ (up to a global phase of $e^{i\pi/4}$) with a PR
by $\pi$ on modes two and three, a PDPS\ of $\pi/2$ on mode two, a PDPS\ of
$-\pi/2$ on mode three.

\section{Probabilistic Controlled-Phase Gate}

The controlled-phase takes each computational basis state to itself except
$\left\vert VV\right\rangle $ becomes $-\left\vert VV\right\rangle $. Suppose
the four computational basis elements are inputs in Figure~\ref{fig_C-Z}.
Suppose the first part of the controlled-phase generates $\left\vert
t_{1}^{\prime}\right\rangle $ (occurring with probability $1/8$). We determine
the propagation of the following four states through the latter half of the
controlled-phase: $\left\vert H\right\rangle _{1}\left\vert t_{1}^{\prime
}\right\rangle _{2345}\left\vert H\right\rangle _{6}$, $\left\vert
H\right\rangle _{1}\left\vert t_{1}^{\prime}\right\rangle _{2345}\left\vert
V\right\rangle _{6}$, $\left\vert V\right\rangle _{1}\left\vert t_{1}^{\prime
}\right\rangle _{2345}\left\vert H\right\rangle _{6}$, $\left\vert
V\right\rangle _{1}\left\vert t_{1}^{\prime}\right\rangle _{2345}\left\vert
V\right\rangle _{6}$. The modes in Figure~\ref{fig_C-Z} are in increasing order
from top to bottom. We determine the state of modes three and four after the
two A2Cs by first analyzing A2C acting on $\left\vert HH\right\rangle $,
$\left\vert HV\right\rangle $, $\left\vert VH\right\rangle $, and $\left\vert
VV\right\rangle $. All four combinations occur when the two A2Cs act on modes
1, 2, 5, and 6 of the above six-mode states.

A2C consists of a beam splitter followed by two PIDs (Figure~\ref{fig_A2C}). A2C
similar to Browne and Rudolph's Type-II fusion gate (though we use a
beamsplitter instead of a polarizing beam splitter) or Pittman et al.'s
\textquotedblleft parity-check gate\textquotedblright, which \textquotedblleft
teleports\textquotedblright\ the input state onto the cluster state with the
effect of performing a logical controlled-phase gate\cite{pra2001franson}.

We describe its operation. Discard the controlled-phase result and start over
if the measurement result is zero photons in exactly three modes. The
operation is a success if zero photons are in exactly two modes. Table
\ref{tbl_A2C post_proc} gives the resulting state and the post-processing.
Table \ref{tbl_A2C post_proc} gives sixteen possibilities each occurring with
probability $1/64$. The controlled-phase success probability is $1/4$ given
$\left\vert t_{1}^{\prime}\right\rangle $.

\section{Conclusion}

We summarize our results. B2G and G2A both have a success probability of
$1/2$. We generate $\left\vert t_{1}^{\prime}\right\rangle $ offline with
success probability $1/8$. Controlled-phase success probability is $1/4$ given
$\left\vert t_{1}^{\prime}\right\rangle $. Photodetector number-resolving
capability is not required throughout all operations---number-resolving
detectors are not necessary for linear optical quantum computation with
cluster states.

We cannot implement our scheme via spontaneous parametric down conversion
without number-resolving detectors. The Bell-state source must be free from
multiple pair emission to purify the Bell states.

M.M.W. thanks Austin Lund for stimulating discussions. We would like to acknowledge
support from the Hearne Institute of Theoretical Physics, the
National Security Agency, the Disruptive Technologies Office, and the Army Research Office. 

\bibliographystyle{unsrt}

\bibliography{cluster-state-LOQC}

\end{document}